\documentclass[12pt]{article}
\usepackage{amsfonts,amsmath}
 \usepackage{amssymb}
 \usepackage[hypertex] {hyperref}

\makeatletter \@addtoreset{equation}{section} \makeatother

\newcommand{\be}{\begin{equation}}
\newcommand{\ee}{\end{equation}}
\newcommand{\bee}{\begin{eqnarray}}
\newcommand{\beee}{\begin{array}}
\newcommand{\eee}{\end{eqnarray}}
\newcommand{\eeee}{\end{array}}

\newcommand{\gee}{\epsilon}

\newcommand{\gf}{\phi}

\newcommand{\ga}{\alpha}
\newcommand{\pa}{{\dot{\ga}}}
\newcommand{\pb}{{\dot{\gb}}}

\newcommand{\gb}{\beta}
\newcommand{\gga}{\gamma}

\newcommand{\M}{{\cal M}}

\newcommand{\ie}{{\it i.e.,} }
\newcommand{\ls}{\!\!\!\!\!\!}

\newcommand{\gd}{\delta}

\newcommand{\gvep}{\varepsilon}

\newcommand{\gs}{\sigma}

\newcommand{\go}{\omega}

\newcommand{\by}{{\bar{y}}}

\newcommand{\q}{\,,\qquad}

\newcommand{\dga}{{\dot{\alpha}}}
\newcommand{\dgb}{{\dot{\beta}}}

\newcommand{\nn}{\nonumber}

\newcommand{\half}{\frac{1}{2}}
\newcommand{\ptl}{\partial}
\newcommand{\p}{\partial}

\newcommand{\f}{\frac}
\newcommand{\C}{{\cal C}}
\newcommand{\R}{{\cal R}}

\newcommand{\U}{\Upsilon}

\begin{document}

\begin{flushright}
{\small FIAN-TD-2014-02}
\end{flushright}
\vspace{1.7 cm}

\begin{center}
{\large\bf Higher-Spin
 Theory and Space-Time Metamorphoses}

\vspace{1 cm}

{\bf  M.A.~Vasiliev}\\
\vspace{0.5 cm} {\it
 I.E. Tamm Department of Theoretical Physics, Lebedev Physical Institute,\\
Leninsky prospect 53, 119991, Moscow, Russia}

\end{center}

\vspace{0.4 cm}

\begin{abstract}
\noindent
Introductory lectures on higher-spin gauge theory given
at 7 Aegean workshop on non-Einstein theories of gravity.
The emphasis is on qualitative features of the higher-spin gauge theory
and peculiarities of its space-time interpretation.
In particular, it is  explained that Riemannian geometry
cannot play a fundamental role in the higher-spin gauge theory. The higher-spin
symmetries  are argued to occur  at ultra high energy scales
beyond the Planck scale. This suggests that the higher-spin
 gauge theory can help to understand Quantum Gravity.
Various types of higher-spin dualities are briefly discussed.
\end{abstract}

\newpage
\tableofcontents

\newpage

\section{Introduction}
\label{intro}

Higher-spin (HS) gauge theories form a class of  theories exhibiting infinite-dimensional symmetries
which go beyond conventional lower-spin symmetries.
The primary goal of these lectures is to focus on qualitative aspects of
HS gauge theories avoiding technical details as much as possible. The emphasis
is on possible consequences of HS symmetries for our understanding of space-time.
It will be explained
in particular that in the setup of HS gauge theories the usual concepts
of Riemannian geometry such as metric, local event
and space-time dimension cannot play a fundamental role. The HS
symmetries  will be argued to occur  at ultra high energy scales
beyond the Planck scale. Having a potential to describe transPlanckian energies,
HS gauge theory can shed light on the problem of  Quantum Gravity.
Various aspects of HS dualities including $AdS/CFT$ and duality with quantum
mechanics are briefly discussed.

\section{Lower-spin global symmetries}

The fundamental example of a lower-spin symmetry is provided by the Poincar\'e
symmetry which underlies relativistic theories. It acts on coordinates of
Minkowski space-time as
$\delta x^a= \gee^a +\gee^{a}{}_{b} x^b$ where
$\gee^a$  and $\gee^{ab}$ are parameters of
infinitesimal translations and  Lorentz rotations, respectively.
One can write
\be
\delta x^a = [T, x^a]\q
T=  \gee^n P_a+\gee^{ab}
M_{ab}\,,
\ee
where
\bee \nn
P_a =\f{\p}{\p x^a}\,, \qquad
M_{ab}=x_a\f{\p}{\p x^b}-x_b\f{\p}{\p x^a}
\eee
are the generators of the Poincar\'e algebra $iso(d-1,1)$  obeying
the commutation relations
\bee\nn
[M_{ab},P_c ]&=&P_a \eta_{bc}-P_b \eta_{ac}\,,
\eee
\be\nn
[M_{ab},\,M_{cd}] = M_{ad}\eta_{bc}-  M_{bd}\eta_{ac}-
 M_{ac}\eta_{bd}+ M_{bc}\eta_{ad}\,,
\ee
\be\nn
[P_a\,, P_b ] = 0\,,
\ee
where $\eta_{ab}$ is the Minkowski metric.

The Poincar\'e algebra admits the (anti-) de Sitter deformation $l$
with
\be\nn
[P_a\,, P_b ] = { \Lambda M_{ab}}\,,
\ee
which describes symmetries of either  anti-de Sitter space at
$\Lambda < 0$ ($l=o(d-1,2)$) or de Sitter space at $\Lambda > 0$ ($l=o(d,1)$). At
$\Lambda = 0$,  $l=iso(d-1,1)$ describes the symmetries of {Minkowski space}.

Supersymmetry is the
 {extension of the Poincar\'e symmetry by} supergenerators $Q_A$
obeying relations
$$\{ Q_A\,,Q_B \} = \gga^a_{AB} P_a\,,
$$

$$[M_{ab}\, , Q_A ] = \gs_{abA}{}^B Q_B \q \gs_{ab} = \f{1}{4} [\gga_a\,,\gga_b ]\,,
$$
where
 $ A, B = 1,2,3,4$ are the Majorana spinor indices in four dimensions.
 Note that, being fermions, supergenerators obey anticommutation relations.

{Internal symmetry}
{generators} $T_i$ {are space-time invariant}
$$
[T_i\,, P_a]=0\q [T^i\,, M_{ab}] =0\,.
$$
{In particular, the symmetries of the Standard Model}
$T_i\in su(3)\times su(2)\times u(1)$ are of this type.

{To complete the list of symmetries that play a role in conventional
lower-spin theories it remains to mention conformal (super)symmetries.
These will be discussed in some more detail below.
}
\section{Local Symmetries}
{A useful viewpoint is that any global symmetry is the remnant of a local
symmetry with parameters like } $ \gvep^a (x), \gvep^{ab}(x), \gvep^\ga(x), \gvep^i(x)$
{ being arbitrary functions of space-time coordinates.}
{ Local symmetries are symmetries of the full theory.}
{ Global symmetries are symmetries of some its particular solution.}

{For example,}
{the infinitesimal diffeomorphisms} $\delta x^a = \gvep^a (x)$ {are symmetries of
GR while the}
{global symmetries with} $\gvep^a (x)= \gee^a + \gee^{a}{}_b x^b$ {are symmetries
of the Minkowski solution} $g_{ab} = \eta_{ab}$ {of the Einstein equations}.

{Let}
$$
S=\int_{M^d} L(\varphi(x), \p_a \varphi (x),\ldots)
$$
{be invariant under a global symmetry} $g$  with parameters
$ \gee^n$ $(n= a,\ga,i,\ldots)$. {Letting the symmetry parameters
be arbitrary functions of space-time coordinates}, $\gee^n\rightarrow \gvep^n(x)$,
we obtain that
$$
\delta S=-\int_{M^d} J_n^a (\varphi) \p_a \gvep^n (x)
$$
since $\delta S$ should be zero at $\p_a \gvep^n (x)=0$.
$J_n^a (\varphi)$ {are conserved currents since} $\p_a J_n^a (\varphi)= 0$
{by virtue of the field equations $\delta S=0$.}

{The local symmetry is achieved with the aid of gauge fields} $A_a^n$
that have the transformation law
$$
\delta A_a^n = \p_a \gvep^n + \ldots\,,
$$
where the ellipsis denotes possible field-dependent terms.
The following modification of the action
$$
S\longrightarrow S+\Delta S+\ldots  \q \Delta S = \int_{M^d} J_n^a (\varphi) A_a ^n (x)
$$
preserves local symmetry in the lowest order in interactions.
The term $\Delta S$ describes the so-called{ Noether current interactions.}

{There is, however, a subtlety}
{if} $\varphi(x)$ {were themselves gauge fields with gauge
parameters} $\gvep^\prime$. In this case it may happen that  $J_n^a (\varphi)$
{is not  invariant under the} $\gvep^\prime$ { symmetry}. Hence
{the Noether current interaction for several gauge fields may be obstructed by gauge
symmetries.}

Localization of various types of lower-spin symmetries leads to important classes
of gauge field theories.

\subsection{Yang-Mills fields}

The Yang-Mills theory is responsible for the localization of internal symmetries.
For a Lie algebra $l$ with generators $T_i$, Yang-Mills fields $A_a^i(x)$
and symmetry parameters $\gvep^i$ are valued in $l$
$$
A_a(x) = A_a^i(x) T_i \q \gvep(x) = \gvep^i (x)T_i\,.
$$
The Yang-Mills gauge transformation is
$$
\delta A_a (x)= D_a \gvep (x)\,,
$$
where
$$
 D_a \gvep (x) = \p_a \gvep (x) + [A_a(x)\,,\gvep (x)]\,
$$
is the covariant derivative.
The commutator of the covariant derivatives gives the Yang-Mills curvature
$$
[D_a\,,D_b] = R_{ab}\q
R_{ab}= \ptl_a A_b -  \ptl_b A_a   + [ A_a , A_b]\,,
$$
which has the transformation law
$$
\gd R_{ab}= [ R_{ab},\gvep]\,.
$$
Needless to say that the Yang-Mills fields play a prominent role in the
modern theory of non-gravitational fundamental interactions, \ie the Standard Model.

\subsection{Einstein-Cartan gravity and supergravity}

Localization of the Poincar\'e symmetry leads to the Cartan formulation of Einstein
gravity. The Yang-Mills gauge fields $A_\nu^n =(e_\nu{}^a, \omega_\nu{}^{ab})$
 associated with the Poincar\'e algebra include the frame field (vielbein)
 $e_\nu{}^a$ and  the Lorentz connection $ \omega_\nu{}^{ab}$. The frame field
$e_\nu{}^a$ {relates base indices} $\nu$ {with the fiber ones} $a$.
(In Minkowski space in Cartesian coordinates, where $e_\nu{}^a =\delta_\nu^a $,
the two types of indices can be {identified}.)
{The gauge transformations have the form}
$$
\delta e_\nu{}^a(x) = \p_\nu \gvep^a(x) +\go_\nu{}^a{}_b(x) \gvep^b(x) - \gvep^a{}_b (x)e_\nu{}^b(x)
+\Delta e_\nu{}^a\,,
$$
$$
\delta \go_\nu{}^{ab}(x) = \p_\nu \gvep^{ab}(x) +
\go_\nu{}^a{}_c(x) \gvep^{cb}(x) - \go_\nu{}^b{}_c(x) \gvep^{ca}(x)
+\Delta \go_\nu{}^{ab}\,.
$$
Here
$ \Delta e_\nu{}^a$ and $\Delta \go_\nu{}^{ab}$ {denote some corrections to
the Yang-Mills transformation law, which are proportional to the curvatures}
$$
R_{\nu\mu}{}^a = \p_\nu e_\mu{}^a +\go_\nu{}^a{}_b e_\mu {}^b - (\nu\leftrightarrow \mu)\q
R_{\nu\mu}{}^{ab} = \p_\nu \go_\mu{}^{ab} +\go_\nu{}^a{}_c \go_\mu {}^{cb} - (\nu\leftrightarrow \mu)\,.
$$
The zero-torsion constraint
$
R_{\nu\mu}{}^a =0$ expresses the Lorentz connection in terms of the frame field
 and its derivatives:
$\omega = \omega (e,\p e)$. In this case $R_{\nu\mu}{}^{\rho\gs}$ equals to
the {Riemann tensor}.
Recall that the relation of the metric with the frame field is
 $g_{\nu\mu}= e_\nu{}^a e_\mu{}^b \eta_{ab}$.

Localization of supersymmetry extends the gravitational fields by
the spin-3/2 gauge field gravitino $\psi_{\nu \ga}$ with the
gauge transformation law
$$\delta \psi_{\nu \ga} = D_\nu \gvep_\ga +\ldots
\,.$$
Gauge theories of this type are called supergravities, constituting a very
interesting class of extensions of the theory of gravity. (See e.g. \cite{Freedman:2012zz} and references therein. Note that the construction of
supergravity in terms of the gauge fields of the supersymmetry algebra was suggested
in \cite{Chamseddine:1976bf}.)

\subsection{Spontaneous symmetry breaking}

Generally, one should  distinguish between the symmetry $G$ of some
equations and a symmetry $\tilde G$ of some their particular solution.
{For example, for the case of the Higgs field} $H^i (x)= H_0^i +h^i(x)$,
{the unbroken part} $\tilde G\subset G$ {is a residual symmetry of} $H_0^i$:
$\tilde G=SU(3)\times U(1)$ {in the Standard Model}.
{For} $H_0^i$ {having a non-zero dimension} $[H_0^i]= cm^{-1}\sim GeV$,
{spontaneous symmetry breaking is a low-energy effect. In other words,
the symmetry restores at} $E> H_0^i$.

{In the unbroken regime, the gauge fields associated with the usual} {lower-spin}
{symmetries describe massless particles} of spin one $ A_\nu{}^i$, spin
$3/2$ $\psi_{\nu\,\ga}$ and
spin two $ e_\nu{}^a$, $ \go_\nu{}^{ab}$.

\section{General Properties of HS Theory}
The key  question is whether it is possible to go to larger {HS} symmetries.
 If yes, what are HS symmetries and
HS counterparts of the lower-spin theories {including GR?}
{What are physical motivations for their study and possible outputs?}

\subsection{Fronsdal fields}
 {As shown by Fronsdal \cite{Fronsdal:1978rb}, all}  symmetric
 massless {HS fields are gauge fields}.
They are described by rank--$s$ symmetric tensors $\gf_{\nu_1\dots \nu_s}$
{obeying} the double tracelessness condition
$\gf^\rho{}_\rho{}^\mu{}_\mu{}_{\nu_5\ldots \nu_s}=0$. The
gauge transformation is
\be
\gd \gf{}_{\nu_1\dots \nu_s}(x)=\p_{(\nu_1}\gvep_{\nu_2\dots \nu_s)}(x)\,,
\ee
where the gauge parameter is symmetric and traceless
\be
{\gvep^\mu{}_\mu{}_{\nu_3\dots \nu_{s-1}}=0}\,.
\ee
The {field equations have the form}
$$
\R_{\nu_1\dots \nu_s}(x)=0
\,,$$
where the {Ricci-like tensor}
$
\R_{\nu_1\dots \nu_s}(x)
$
is
$$
\R_{\nu_1\dots \nu_s}(x)=\Box \gf_{\nu_1\dots \nu_s}(x)-
s\p_{(\nu_1}\p^\mu\gf_{\nu_2\dots \nu_{s}\mu)}(x)
+\f{s(s-1)}{2}\p_{(\nu_1}\p_{\nu_2}\gf^\mu{}_{\nu_3\dots \nu_{s}\mu)}(x)\,.
$$
{The gauge invariant Fronsdal action is}
$$ \nn S=
\int_{M^d} \Big(\half \gf^{ \nu_1\dots \nu_{s}}\R_{ \nu_1\dots \nu_{s}}(\gf)
-\f{1}{8}s(s-1)\gf_\mu{}^{\mu\, \nu_3\dots \nu_{s}}
\R^\rho{}_{\rho\, \nu_3\dots \nu_{s}}(\gf)\Big )\,.
$$

\subsection{No-go and the role of $(A)dS$}
{In the sixties of the last century it was argued by Weinberg \cite{Weinberg:1965rz} and
Coleman and Mandula \cite{Coleman:1967ad} that  HS symmetries
cannot be realized  in a nontrivial local field theory in Minkowski space.
In the seventies it was shown by Aragone and Deser \cite{Aragone:1979hx}
that HS gauge symmetries
are incompatible with GR within an expansion over  Minkowski space}. The general
belief
was that nontrivial interactions of massless HS fields cannot be introduced.

Nevertheless, in the eighties, it was shown by light-cone
\cite{Bengtsson:1983pd,Bengtsson:1983pg} and covariant methods
\cite{Berends:1984wp,Berends:1984rq} that some non-gravitational HS
interactions can be constructed at least at the cubic order.
These results suggested that some consistent HS theory should exist.

The further progress resulted from the observation that the consistent
formulation of the HS gauge theory requires a curved background instead
of the flat Minkowski. The most symmetric curved cousins of the flat Minkowski
space are de Sitter and anti-de Sitter spaces. That HS theories admit
consistent interactions including the gravitational interaction in $(A)dS$
 background was shown in \cite{Fradkin:1987ks,Fradkin:1986qy}.
{In agreement with the no-go statements, the limit} of zero cosmological constant
$\Lambda\to 0$ {turns out to be
 singular} so that, indeed, HS theories with unbroken HS symmetries do not exist in
the Minkowski background.

\subsection{HS Symmetries versus Riemannian  geometry}

{The HS symmetries and the space-time symmetries do not commute
simply because HS generators are higher-rank Lorentz tensors}
$$
[T^a\,, T^{HS}] = T^{HS}\q [T^{ab}\,, T^{HS}] = T^{HS}\,.
$$
However, the same commutation relations imply that HS generators transform
the space-time generators to the HS generators. Since the gauge fields for space-time
generators are the gravitational frame field and Lorentz connection, this implies
 that HS transformations map the gravitational fields (metric) to the HS fields.

This simple observation has the far-going consequence that
the Riemannian geometry is not appropriate for the HS theory,
implying in particular that the concept of
local event may become illusive in the HS theory!

Though it is not appropriate to use the metric tensor in the HS theory,
we do not want to give up the coordinate independence of GR. Fortunately,
this can be achieved in the framework of the formalism of differential forms.

{Differential forms are totally antisymmetric tensors.}
A $p$-{form} is a rank-$p$ totally antisymmetric
tensor $\go(x)=\theta^{\nu_1}\ldots \theta^{\nu_p}\go_{\nu_1\ldots \nu_p}(x)$
where $\theta^\nu$ are anticommuting symbols (variables)
$$
\theta^\nu \theta^\mu = -\theta^\mu \theta^\nu
$$
usually called differentials $\theta^\nu = dx^\nu$.
{The invariant differentiation is provided by the exterior (de Rham) derivative}
$$
d=\theta^\nu \f{\p}{\p x^\nu}\q d^2 =0\,.
$$
{This formalism is covariant because, due to the  total
antisymmetrization of indices, symmetric Christoffel symbols drop out from
the covariant derivatives.} In this language, the
{connections} $A= \theta^\nu A_\nu^i T_i$ {are} {one-forms}, while the
{curvatures} $R=D^2$ with $D=d+A$ {are} {two-forms.}

{Farther elaboration of this language in application to
HS theory leads eventually to a deeper understanding
of fundamental concepts of space-time including its dimension.}

\subsection{HS gauge theory, Quantum Gravity and String Theory}
{As explained in more detail below, the HS symmetry is in a certain sense
{maximal} relativistic symmetry. Hence one can speculate that
it cannot result from spontaneous breakdown of a larger symmetry}.
This implies that
{the HS symmetries are manifest at ultrahigh energies above any scale including
the Planck scale.} If this is true,
 {the HS gauge theory should capture effects of Quantum Gravity}.
 This opens a unique possibility for the theoretical study of
the unreachable by experimental
tests energy scale of Quantum Gravity by means of the highly restrictive HS symmetry.

On the other hand, since the lower-spin symmetries are subalgebras of
the HS symmetries, it is natural to expect that the
lower-spin theories  can correspond to low-energy limits
of the HS theory with spontaneously broken HS symmetries.

A related issue is a connection of HS theory with String Theory.
A natural conjecture is that String Theory can be interpreted
as a spontaneously broken theory of the HS type, where  $s>2$ fields
acquire nonzero masses. An interesting recent conjecture
\cite{Chang:2012kt} is that
 String Theory can be identified with the full quantum HS theory.

\subsection{Higher-spin $AdS/CFT$ correspondence}

That the HS gauge theories are most naturally formulated in the anti-de Sitter
background makes them interesting from the perspective of $AdS/CFT$ correspondence
\cite{Maldacena:1997re,Gubser:1998bc,Witten:1998qj}. Various aspects of
the HS holography were discussed
by many authors starting from
 \cite{Konstein:2000bi,Sundborg:2000wp,WJ} (see also
\cite{Mikhailov:2002bp,Sezgin:2002rt}).
However, the concrete proposal
is due to Klebanov and Polyakov \cite{Klebanov:2002ja} who conjectured that
the $AdS_4$  {HS theory  is dual to} $3d$ {the vectorial conformal
 models. This hypothesis was successfully checked by Giombi and Yin
 \cite{Giombi:2009wh}, that triggered a lot of interest to the HS holography.
The conjecture of Klebanov and Polyakov was later extended to the
fermionic boundary systems \cite{Leigh:2003gk,Sezgin:2003pt} as well to
the  $AdS_3 /CFT_2 $ {correspondence}
\cite{Henneaux:2010xg,Campoleoni:2010zq,Gaberdiel:2010pz}.

The HS holography has several features which give a hope that its analysis
may help to uncover the origin of $AdS/CFT$. Indeed, as discussed in some
more detail below, a progress in this direction has been achieved in
\cite{Vasiliev:2012vf}.
It should be stressed that the  HS holography does not rely on supersymmetry and
is a weak-weak duality that therefore can be checked directly on the both sides.
For more detail on the HS holography we refer the reader to
\cite{Gaberdiel:2012uj,Giombi:2012ms}.

\section{Global HS Symmetry: Idea of Construction}
The simplest way to figure out what is a HS symmetry is via the
$AdS/CFT$ correspondence. Namely, the global
{HS symmetry} of the most symmetric  $AdS_{d+1}$ solution
{can be identified with the maximal symmetry of the} $d$-dimensional
free conformal fields. In the most cases the latter are identified with the
massless scalar  and/or spinor.

Consider KG massless equation in $d$-dimensional Minkowski space
$$
\Box C(x) = 0\q \Box = \eta^{ab}\f{\p^2}{\p x^a \p x^b}\,.
$$
The conformal HS symmetry is the symmetry of this equation.
{What is this symmetry?} Its structure was first elaborated
for $d=3$ in \cite{Shaynkman:2001ip} and soon after
{by Eastwood \cite{Eastwood:2002su} for any $ d$}.

Of course, this symmetry contains the {Poincar{\'e}} transformations
as well as {the scale transformation (dilatation)}
$$
\delta C(x) = \gvep DC(x)\q D = x^a \f{\p}{\p x^a} +\f{d}{2} -1
$$
{and the special conformal transformations}
$$
\delta C(x) = \gvep_a K^a C(x)\q K^a = (x^2\eta^{ab} -2 x^a x^b )\f{\p}{\p x^b} +(2-d) x^a\,.
$$
Altogether $P_a, M_{ab}, K^a$ and $D$ form the conformal Lie algebra
$o(d,2)$.

To figure out the structure of the whole conformal HS algebra it is
useful to consider an auxiliary problem.

\subsection{Auxiliary problem}
{Consider the equations}
\be
\label{1}
D \C_A(x) =0\,,
\ee
where $\C_A(x)$ is a set of fields valued in some space $V$ (the label $A$)
and
$$
D=d+\go(x)\q \go_A{}^B(x) = \go^\Omega(x) T_{\Omega\,A}{}^B
$$
is a covariant derivative acting in the space $V$ treated as a
$gl(V)$-module. \ie $\go(x)$ is some $gl(V)$-connection. The covariant
derivative $D$ is demanded to be flat, \ie
\be
\label{flat}
D^2 =0\,.
\ee
{Clearly, Eqs.~(\ref{1}) and (\ref{flat}) are
invariant under the gauge transformation}
$$
\delta \C_A(x) = - \gvep_A{}^B(x) \C_B(x)\q  \gvep_A{}^B(x) = \gvep^\Omega (x)
T_{\Omega\,A}{}^B (x)\,,
$$
$$\delta \go(x) = D\gvep(x):= d\gvep(x) +\go(x) \gvep(x)
- \gvep(x) \go(x)\,,
$$
where indices are implicit. {The condition that the equations remain invariant
{for some fixed} $\go(x)=\go_0(x)$ restricts the
gauge parameters $\gvep^\Omega(x)$ to the  parameters
$ \gvep_{gl}^\Omega(x)$ obeying the  conditions
$$
\delta \go_0(x)=0\quad \longrightarrow \quad D_0 \gvep_{gl}^\Omega(x) =0
\q D_0 := d +\go_0\,.
$$
{Since} $D^2_0 =0$,  $\gvep_{gl}^\Omega(x)$ {is reconstructed (locally)
in terms of}
$\gvep_{gl}^\Omega(x_0)$ at any $ x_0$. $\gvep_{gl}^\Omega(x_0)$
{are the global symmetry parameters of the equation} $D_0 \C(x) =0$.

{Alternatively, one can write a solution in the pure gauge form}
$$\go_0(x) = g^{-1}(x) d g(x)\q \C(x) = g^{-1} (x) \C\q\gvep_{gl}(x) =
g^{-1}(x) \gee g(x)\,.
$$
{For} $g(x_0)=1$ this gives $\C= \C(x_0)$ and $\gee = \gvep_{gl}(x_0)$.

\subsection{Massless scalar field unfolded}
\label{massunf}
{Minkowski space is described by a flat Poincar\'e-connection}
$\go(x) = e^a(x) P_a +\half \go^{ab}(x) M_{ab}$. In
 {Cartesian coordinates} $ e^a(x) = \theta^a$ and $\go^{ab}=0$.

{Introduce an infinite set of zero-forms, which are traceless symmetric tensors}
\be
\label{scal}
 C_{a_1\ldots a_n}(x)=C_{(a_1\ldots
a_n)}(x)\,,\quad \eta^{bc}C_{bca_3\ldots a_n}(x)=0\,.
\ee
{The unfolded system of equations equivalent to the Klein-Gordon equation
has the form}
\be\label{un0}
\displaystyle d C_{a_1\ldots
a_n }(x) =\theta^b C_{a_1 \ldots a_n b}(x) \,.
\ee
Since the fields $C_{a_1\ldots a_n}(x)$ are symmetric while
$\theta^b \wedge \theta^c=-\theta^c \wedge \theta^b$,
{the system (\ref{un0}) is formally consistent.
(Equivalently, the covariant derivative associated with the equation
(\ref{un0}) rewritten in the form (\ref{1}) is flat.)

{The first two equations imply}
$$\partial_a C(x) =C_a (x) \,,\qquad
\partial_a C_b(x)= C_{ab}(x)\longrightarrow C_{ab}(x)=\partial_a\partial_b C(x)\,.
$$
{Since $C_{ab}(x)$ is traceless} this implies
\be
\label{KG}
\Box C(x) =0\,.
\ee
{ All other equations express higher tensor components via higher derivatives of
the scalar field}
\be
\label{cons}
 C_{a_1 \ldots a_n}(x)=
\partial_{a_1}\ldots\partial_{a_n}C(x)\,.
\ee
This formula explains the meaning of
$C_{a_1 \ldots a_n}(x)$ as spanning a basis of the space of {all on-mass-shell nontrivial
derivatives of} $C(x)$. It should be noted that the space of $C_{a_1 \ldots a_n}(x)$  is
 analogous (in some sense dual) to the space of single-particle states.
 Via Eq.~(\ref{un0}) the set of fields $ C_{a_1 \ldots a_n}(x)$ at any given
 $x=x_0$ determines $C(x)$ in some neighborhood of $x_0$, thus providing a
 locally complete set of ``initial data".

\subsection{Any $d$}
{}From the unfolded form of the massless scalar field equations it follows
that {the conformal HS algebra $h$ in} $d$ {dimensions is the algebra of linear transformations
of the infinite-dimensional space} $V$ {of various traceless symmetric tensors}
$C, C_a, C_{ab}\ldots$, \ie {$h=gl(V)$}. Since the space $V$ is infinite dimensional,
such a definition is not fully satisfactory, requiring a more precise
definition of the appropriate class of operators.
In practice, the idea is that the basis operators of the  conformal HS algebra
$h$ should reproduce the HS symmetry transformations represented by
finite-order differential operators.

A careful definition of
$h$ {was given by Eastwood in \cite{Eastwood:2002su} by different methods}.
As shown in \cite{Shaynkman:2001ip}, the construction for $d=3$
significantly simplifies in the framework of the spinorial formalism. Since this
formulation is most relevant
in the context of the $AdS_4/CFT_3$ {HS holography} we explain it in some more
detail.

\section{Conformal HS Algebra in $d=3$}
\label{Conformal HS}
\subsection{$3d$ multispinors}
Convenience of the language of spinors in $3d$ theories is due to the
following well-known isomorphisms of the
$3d$ {Lorentz algebra:} $o(2,1)\sim sp(2,R)\sim sl_2(R)$. $3d$ {spinors
in Minkowski signature are real}
$$
\chi^\dagger_\ga = \chi_\ga\q \ga=1,2\,.
$$
The $sp(2,R)$ {invariant tensor} $\epsilon^{\ga\gb} = - \epsilon^{\gb\ga}$
{relates lower and upper indices}
$$
\chi^\ga = \epsilon^{\ga\gb} \chi_\gb\q \chi_\ga = \chi^\gb \epsilon_{\gb\ga}\,.
$$
{Because a two-by-two antisymmetric matrix is unique up to a factor, the
antisymmetrization of $3d$ spinor indices is equivalent to their contraction}
$$
A_{\ga,\gb} - A_{\gb,\ga} = \epsilon_{\ga\gb} A_{\gga,}{}^\gga\,.
$$
{As a result,  irreducible modules of the Lorentz algebra are represented by various totally symmetric
multispinors} $A_{\ga_1\ldots \ga_n}$.
{As a  consequence, rank-$k$ traceless symmetric tensors in the
tensor notations
are equivalent to the rank-$2k$ totally symmetric multispinors}
$$
A_{a_1\ldots a_m}\sim A_{\ga_1\ldots \ga_{2m}}\q A^b{}_{b a_3\ldots a_m}=0\,.
$$
(The reader can compare the number of independent components of the both
objects).

{The explicit relation between  the two formalisms is established
with the help of the} $2\times 2$ {real symmetric matrices} $ \sigma_{\ga\gb}^n$
$$
A_{\ga\gb} = \sigma_{\ga\gb}^n A_n \q \sigma_{\ga\gb}^n = \sigma_{\gb\ga}^n\,.
$$

\subsection{Spinorial form of $3d$ massless equations}
{ In $d=3$, the space} $V$ {of all}   {traceless symmetric tensors is
equivalent to
the space of even functions of the commuting spinor variable} $y^\ga$
$$
C(y|x) = \sum_{n=0}^\infty C^{\ga_1\ldots \ga_{2n}}(x) y_{\ga_1}\ldots y_{\ga_{2n}}\,.
$$
{In these terms, the unfolded equations for a massless scalar take the form}
\be\label{2}
\theta^{\ga\gb} \left ( \f{\p}{\p x^{\ga\gb}} + \f{\p^2}{\p y^\ga \p y^\gb}\right ) C(y|x) =0
\ee
with $C(-y|x) = C(y|x)$.
The same equation with  odd $C(-y|x) = - C(y|x)$  describes a $3d$
massless spinor field $C_\ga(x)= \f{\p}{\p y^\ga} C(y|x)\Big |_{y=0}$
\cite{Shaynkman:2001ip}.

\subsection{$3d$ HS symmetry}

The $3d$ bosonic {conformal HS algebra is the algebra of various
differential operators} $\epsilon(y,\f{\p}{\p y} )$ {obeying}
$$
\epsilon (-y,-\f{\p}{\p y}) = \epsilon (y,\f{\p}{\p y})\,.
$$
The transformation law is
\be
\label{dc}
\delta C(y|x) = \gvep_{gl} (y,\f{\p}{\p y}|x) C(y|x)\,,
\ee
where
\be
\label{epsgl}
\gvep_{gl} (y,\f{\p}{\p y}|x) = \exp{\left [-x^{\ga\gb} \f{\p^2}{\p y^\ga \p y^\gb}\right ]}
\epsilon (y,\f{\p}{\p y}) \exp{\left [x^{\ga\gb} \f{\p^2}{\p y^\ga \p y^\gb}\right ]}\,.
\ee
We leave it to the reader to check that this transformation indeed maps a solution
of (\ref{2}) to a solution.
{For any polynomial } $\epsilon (y,\f{\p}{\p y})$, $\epsilon_{gl} (y,\f{\p}{\p y}|x) $
is polynomial as well.
 $\epsilon_{gl} (y,\f{\p}{\p y})$ {provides the generating function for
parameters of the global HS transformations.}

The $3d$  conformal algebra $sp(4)\sim o(3,2)$ is a subalgebra of the HS conformal
algebra with the generators
\be
\label{bilin}
P_{\ga\gb} = \f{\p^2}{\p y^\ga \p y^\gb}\q K^{\ga\gb} = y^\ga y^\gb\q
M_{\ga\gb} = y_\ga\f{\p}{\p y^\gb} + y_\gb\f{\p}{\p y^\ga}\q
D=y^\ga \f{\p}{\p y^\ga} +1\,.
\ee
It is not difficult to check how formula (\ref{dc}) reproduces the standard conformal
transformations for massless scalar and spinor in three dimensions.

\subsection{Weyl algebra and star product}
{The Weyl algebra} $A_n$ {is the associative  algebra of polynomials of oscillators}
$\hat Y_A$ obeying the commutation relations
\be
\label{YY}
[\hat Y_A\,, \hat Y_B] = 2i C_{AB}\q A,B,\ldots = 1,\ldots 2n \q C_{AB}=-C_{BA}
\ee
with a nondegenerate $C_{AB}$. Taking into account that
$$
\hat Y_A = \left ( {y^\ga}\atop{i\f{\p}{\p y^\gb}}\right )
$$
obey the Heisenberg commutation relations (\ref{YY}), we conclude that the
$3d$ {conformal HS algebra (to be identified with the $AdS_4$ HS algebra)
is the Lie algebra associated with the even part of the Weyl algebra} $A_2$.

In practice, it is convenient to replace any operator
$$
\hat f(\hat Y)= \sum_{n=0}^\infty \f{1}{n!}f^{A_1\ldots A_n}
\hat Y_{A_1}\ldots \hat Y_{A_n}$$
with symmetric $f^{A_1\ldots A_n}$ by its Weyl
{symbol} $f({Y})$ which is the function of commuting
variables $Y^A$ ($Y^A Y^B = Y^B Y^A$), {that has the same power series expansion}
$$
 f( Y)= \sum_{n=0}^\infty \f{1}{n!}f^{A_1\ldots A_n}  Y_{A_1}\ldots  Y_{A_n}\,.
$$

{The Weyl star product  is defined by the  rule that}
$
(f*g) (Y)$  is the symbol of  $\hat{f}(\hat Y)
\hat{g}({\hat Y})\,.
$
 In particular, this implies
$$
\label{bstar} [{Y}_A,Y_B]_*=2iC_{AB}\q
[a\,,b]_* = a*b -b*a\,.
$$
One can also see that
$$
\{Y_A\,,f(Y)\}_*= 2Y_A f(Y)\q [Y_A\,,f(Y)]_* = 2i\f{\p}{\p Y^A} f(Y)\,,
$$
where
$$
Y^A =C^{A B}Y_B\,.
$$

The star product is concisely described by the {Weyl-Moyal formula}
\be
\label{moyal}
(f_1*f_2)(Y)=f_1(Y)\,
\exp{[i\overleftarrow{\partial^A}\overrightarrow{\partial^B}
C_{AB}]} \,f_2(Y) \ ,\quad
\partial^A:=\frac{\partial}{\partial Y_A}\,,
\ee
which can be proven using the Campbell-Hausdorff formula for $\exp{J^A \hat{Y}_A}$.

{By its definition, the star product (\ref{moyal}) is associative}
 $(f*g)*h =f*(g*h)$ {and  regular in the sense that the}
{star product of any two polynomials of} $Y$ { is
a polynomial}.

The star product also admits the following useful
{ integral representation}
\bee \nn\label{trig} (f_1 *
f_2)(Y)=\frac{1}{\pi^{2M}}\int dS dT &{}&\ls\exp(-iS_A T_B C^{AB})
 f_1(Y+S)\,f_2(Y+T)\,.
\eee

\section{HS Symmetry in $AdS_4$}

\subsection{Spinor language in four dimension}
The HS theory in four dimensions is most naturally formulated in terms of
two-component spinors which language is closely related to the twistor theory.
Here {the  key fact is that} $2\times2=4$.
{ Minkowski coordinates are represented by} $2\times2$ { Hermitian matrices}
$$
X^n\Rightarrow X^{\ga\pa} =\sum^3_{n=0}X^n \gs_n^{\ga\pa}\q
\gs_n^{\ga\pa}= (I^{\ga\pa},\vec{ \gs}^{\ga\pa})\,,
$$
where
$I^{\ga\pa}$ is the unit matrix and
$\vec{ \gs}^{\ga\pa}$  {are Pauli matrices}.
$\ga,\gb,\ldots =1,2$, $\pa,\pb,\ldots =1,2$
{are two-component spinor indices}.

In these terms
$$
\det|X^{\ga\pa} | = (X^0)^2 - (X^1)^2 -(X^2)^2 -(X^3)^2\,.
$$
This  relation establishes the well-known isomorphism for
the four-dimensional
{Lorentz algebra} $sl(2,{\mathbb C})\sim o(3,1)$.

{The dictionary between tensors and multispinors is provided by the $\sigma$-matrices}
$$
\sigma^a_{\ga\dga}\q \sigma^{ab}_{\ga\gb}=\sigma^{[a}_{\ga\dga}\sigma^{b]}_{\gb}
{}^\dga\q
\bar{\sigma}^{ab}_{\dga\dgb}={\sigma}^{[a}_{\ga\dga}\sigma^{b]\ga}{}_\dgb\,,
$$
where the two-component indices are raised and lowered by the two-by-two
antisymmetric form $\gvep_{\ga\gb}$,
\be\nn
y^\ga = \gvep^{\ga\gb}y_\gb\q y_\ga = y^\gb \gvep_{\gb\ga}\q
\gvep_{\ga\gga}\gvep^{\gb\gga}=\delta_\ga^\gb\q
\gvep_{12}=\gvep^{12}=1\,.
\ee
These relations show that
{a pair of dotted and undotted indices is equivalent to a  vector index,}
{while the pairs of symmetrized indices of the same type are equivalent to
the second-rank antisymmetric tensors}.

\subsection{$AdS_4$ HS algebra}
{The identification of the $3d$ conformal HS symmetry with the $AdS_4$ HS symmetry}
implies that
{the global symmetry of the most symmetric  vacuum  of the bosonic HS theory
  is represented by the Lie algebra associated with the even part of the
  Weyl algebra $A_2$. To have $4d$ Lorentz symmetry manifest, it is most convenient
  to realize $A_2$ as the algebra of mutually conjugate operators $y_\ga$ and
  $ \bar y_\dga$ that obey the star-product commutation relations
$$
[y_\ga\,,y_\gb ]_*=2i\gvep_{\ga\gb}\q
[\bar y_\dga\,,\bar y_\dgb ]_*=2i\gvep_{\dga\dgb}\,.
$$
Historically, the $AdS_4$ HS algebra was originally found in \cite{Fradkin:1986ka}
by different methods from the analysis
of the HS fields in $AdS_4$ while its relation to the
Weyl algebra was found in \cite{Vasiliev:1986qx}.

This realization is convenient for the analysis of the properties of the HS algebra.}
{Spin-$s$ generators are represented by the homogeneous
polynomials} $T_{s}(y,\bar y)$  {of degree} $2(s -1)$. The commutation relations
have the following structure
 \bee\nn
\label{ee} [T_{s_1} \,,T_{s_2} ]=T_{s_1+s_2-2}+T_{s_1+s_2-4}+\ldots +T_{|s_1-s_2|+2}\nn
\,.
\eee
Once a spin $s>2$ {appears, the HS  algebra contains an infinite tower of
higher spins. Indeed,} since $[T_s,T_s]$ {gives rise to} $T_{2s-2}$, further
commutators then lead to higher and higher spins. {Note also that $[T_s,T_s]$}
contains the generators $T_2$ {of} the $AdS_4$ algebra $o(3,2)\sim sp(4)$.

{The HS gauge fields in four dimensions are} the one-forms
$$
\go(Y|X)=\sum_{n,m=0}^\infty\f{1}{2 n!m!}
\go_{\ga_1\ldots \ga_n\,,\dga_1\ldots \dga_m}(X)y^{\ga_1}\ldots y^{\ga_n}
\bar{y}^{\dga_1}\ldots \bar{y}^{\dga_m}\,,
$$
where  $Y_A=(y_\ga, \bar y_\dga)$ are commuting spinor variables and $X$ are
local coordinates of $AdS_4$.
The  HS curvatures and gauge transformations are
\be
\label{curv}
R(Y|X)= d\go(Y|X) +\go(Y|X)* \go(Y|X)\,,
\ee
\be
\label{gtr}
\delta \go(Y|X) = D\epsilon(Y|X)= d\epsilon(Y|X)+ [\go(Y|X)\,,\epsilon(Y|X)]_*\,.
\ee

The symmetry algebra of  a single boundary scalar field called $hu(1,0|4)$
contains every spin in one copy.
Conventional symmetries are associated with  spins  $s\leq 2$,
forming finite-dimensional subalgebras of the HS algebra. For example,
 the maximal
finite-dimensional subalgebra of $hu(1,0|4)$ is $u(1)\oplus o(3,2)$ where
$u(1)$  is  associated with the unit element of the star-product algebra.

More generally, there are
{three series of} $4d$ {HS superalgebras}, namely $hu(n,m|4)$, $ho(n,m|4)$
and  $husp(2n,2m|4)$.
{Spin-one fields of the respective HS theories are
the Yang-Mills fields of the Lie groups} $G=U(n)\times U(m)$, $O(n)\times O(m)$ {and}
$Usp(2n)\times Usp(2m)$, respectively.
{Fermions belong to the  bifundamental modules of the two components of $G$.}
{All odd spins are in the adjoint representation of } $G$.
 Even spins carry  the opposite symmetry second rank
representation of $G$. Namely, in the $hu(n,m|4)$ HS theories they are still
in the adjoint representation of $U(n)\times U(m)$, while in the
$ho(n,m|4)$
and  $husp(2n,2m|4)$ HS theories even spins carry rank-two symmetric
representation of $O(n)\times O(m)$ and antisymmetric representation of
$Usp(2n)\times Usp(2m)$, respectively. The $ho(1,0|4)$ HS theory is  the
minimal one only containing  even spins $s=0,2,4,6,\ldots$.

The HS theories have the important feature that their particle spectrum always
contains a
 {colorless graviton} and {a colorless scalar} which are both invariant under the
 spin-one Yang-Mills
internal symmetries. It is interesting to note that the presence of the colorless
scalar field in the spectrum, which is a standard ingredient of the modern
cosmological models, is one of the predictions of the HS symmetry.

\section{Free HS fields in four dimension}
\subsection{Vacuum solution}

Whatever form they have, nonlinear HS field equations will be formulated in
terms of the HS
curvatures. Hence, any connection $\go(Y|X)$ that has zero curvature
solves the nonlinear  HS equations of motion. Such solutions include
in particular the $AdS_4$  connection
because $AdS_4$ is described by the flat gravitational connections of $sp(4)$
which is a subalgebra of the HS algebra.

The $AdS_4$ {vacuum solution solves the equations}
$$
R_0=0 $$
 { for} $\go_0\in sp(4)\sim o(3,2)$ that has the form
$$
\go_0(Y|X)=\f{1}{4i} (w^{\ga\gb}(X)y_\ga y_\gb+ \bar w^{\dga\dgb}(X)
\bar{y}_\dga\bar{y}_\dgb  +2\lambda  h^{\ga\dgb}(X) y_\ga \bar{y}_\dgb )\,.
$$
{We leave it to the reader to check that these equations indeed describe $AdS_4$.

{Fluctuations describe small deviations of all massless fields from the vacuum}
$$
\go=\go_0 +\go_1\q
 R_1=D_0\go_1:= d\go_1 +[\go_0\,, \go_1]_*\,.
$$
Since we know free massless field equations, we anticipate them to result from the
linearization of the full nonlinear system. The key question is in which
 form the free massless field equations  will follow from the full nonlinear
 system?  The appropriate form is provided by the Central on-shell theorem.

\subsubsection{Central on-shell theorem}
{The full unfolded system for the free massless fields of all spins can be
 formulated in terms of} the
{one-form} $\go(Y|X)$ {and zero-form} $C(Y|X)$ as follows \cite{Vasiliev:1988sa}:
\be\label{eqr}
R_1(Y\mid X) =  \overline{H}^{\dga\pb}
\f{\p^2}{\p \overline{y}^{\dga} \p \overline{y}^{\dgb}}
C(0,\overline{y}\mid X) + H^{\ga\gb}
\f{\p^2}{\p {y}^{\ga} \p {y}^{\gb}}C(y,0\mid X) \ ,
\ee
\be
\label{eqc}
\tilde{D}_0C(Y\mid X) =0\,,
\ee
{ where}
$$
H^{\ga\gb} = h^{\ga}{}_\dga \wedge h^\gb{}^\dga\,,\quad
\overline {H}^{\dga\pb} = h_{\ga}{}^\dga\wedge h^{\ga\pb}
$$
are the basis two-forms in four dimension,
$$
R_1 (Y\mid X) =D_0^{ad}\omega (Y\mid X)\,,
$$
$$
D_0^{ad} = D^L  -
\lambda h^{\ga\pb}\left(y_\ga \frac{\partial}{\partial \bar{y}^\pb}
+ \frac{\partial}{\partial {y}^\ga}\bar{y}_\pb\right) \q
\tilde{D}_0  = D^L  +\lambda h^{\ga\pb}
\left(y_\ga \bar{y}_\pb +\frac{\partial^2}{\partial y^\ga
\partial \bar{y}^\pb}\right)\,,
$$
and the Lorentz covariant derivative is
$$
D^L A  = d_X  -
\left(\go^{\ga\gb}y_\ga \frac{\partial}{\partial {y}^\gb} +
\bar{\go}^{\dga\pb}\bar{y}_\dga \frac{\partial}{\partial \bar{y}^\pb} \right)\,.
$$

Since the system of equations (\ref{eqr}) and (\ref{eqc})
 contains the exhaustive  information about free
 massless fields, including all their dual formulations, it is  called
 Central on-shell theorem.

The pattern of Eqs.~(\ref{eqr}) and (\ref{eqc}) is as follows.
{The gauge fields of different spins are described by the homogeneous polynomials in} $Y$
$$
\go^s(\nu y, \nu \bar y|X) = \nu^{2(s-1)} \go(y,\bar y|X)\,.
$$
The zero-forms associated with the spin $s$ obey
$$
C^s(\nu y, \nu^{-1} \bar y|X) = \nu^{\pm 2s} C(y,\bar y|X)\,.
$$
This implies that a set of one-forms associated with  a massless
spin $s$ contains a finite number of components while a set of zero-forms
contains an infinite number of components. Altogether, these fields describe an
{infinite set of spins} $s=0,1/2,1,3/2,2,5/2\ldots $
\be
\label{list}
\go^s_{\ga_1\ldots \ga_n\,,\dgb_1\ldots \dgb_m}:\quad n+m=2(s-1)\q
C^s_{\ga_1\ldots \ga_n\,,\dgb_1\ldots \dgb_m}:\quad |n-m|=2s\,.
\ee

The zero-forms $C(Y|X)$ {encode the gauge invariant HS curvatures and spin-zero matter fields
along with all their derivatives that remain non-zero on the dynamical field
equations.} Dynamical fields include the frame-like fields
$
\go^s_{\ga_1\ldots \ga_{s-1}\,,\dgb_1\ldots \dgb_{s-1}}$
and the scalar $C(0,0|x)$. The frame-like fields reduce to the Fronsdal fields
upon gauge fixing of the Lorentz-like Stueckelberg gauge symmetries in the
linearized HS gauge transformation (\ref{gtr}).

All other fields from the list (\ref{list}) are expressed by
Eqs.~(\ref{eqr}) and (\ref{eqc}) via higher derivatives of the dynamical fields.
The derivatives come in the dimensionless combination
$$
\lambda^{-1} \f{\p}{\p x}\q \lambda^2=-\Lambda\,
$$
with the inverse  radius
$\lambda$ of the background $AdS$ space.
As a result, the HS interactions, that contain higher derivatives, turn out to
be nonanalytic in} the cosmological constant $\Lambda$ of the background
$AdS$ space.

\subsubsection{Examples}

In the spin-zero sector, the Central on-shell theorem just reproduces the unfolded
equations for a scalar field. Indeed, the set of all multispinors
$ C^0_{\ga_1\ldots \ga_n\,,\dgb_1\ldots \dgb_n}$ with the equal numbers of
dotted and undotted spinor indices provides
the spinorial realization of the set of all symmetric traceless tensors
$C_{a_1\ldots a_n}$, $ C^b{}_{ba_3\ldots a_n}=0$ in four dimension.

Leaving the derivation of the Maxwell equations in the spin-one sector
to the reader,
we consider the case of spin two. Here the
gauge fields include the Lorentz connection $\go_{\ga\gb}, \bar \go_{\dga\dgb}$
 and the frame field  $\go_{\ga,\dgb}$.
{The zero-forms} $C_{\ga_1\ga_2\ga_3\ga_4}(X)$ {and}
$\bar C_{\dga_1\dga_2\dga_3\dga_4}(X)$ {describe the Weyl tensor in
terms of two-component spinors. Higher components}
$C^s_{\ga_1\ldots \ga_n\,,\dgb_1\ldots \dgb_m}$ with
 $|n-m|=4$ {describe all its non-trivial derivatives}.

Consider first Eq.~(\ref{eqr}).
The equation $R_{\ga,\dgb}=0$ {is the usual zero-torsion condition
that expresses the Lorentz connection via the vierbein.} The other equations
have the form
\be
R_{\ga\gb}= H^{\gga\delta}C_{\ga\gb\gga\delta}\q
 R_{\dga\dgb}=\bar
H^{\dot{\gga}\dot{\delta}}\bar C_{\dga\dgb\dot{\gga}\dot{\delta}}\,.
\ee
{These imply that a nonzero part of the Riemann tensor
belongs to the Weyl tensor.} This is equivalent to saying that
the Ricci tensor is zero which, in turn, is equivalent to the
Einstein equations in the vacuum.

In the tensorial language the same equations read as
\be
\label{sys2}
R_{\nu\mu}{}^a =0\q R_{\nu\mu}{}^{ab} = e_{\nu}{}^c e_\mu{}^d
C_{cd,}{}^{ab}{}\q C_{ab,}{}^b{}_c =0\,.
\ee
This implies the Einstein equations since
$R_{\nu\mu} = R_{\nu \rho}{}^\rho{}_\mu =0$.
In addition, the system (\ref{sys2}) implies that
$C_{cd,}{}^{ab}$ coincides with the Weyl tensor.

Analogously,  the Central on-shell theorem for higher spins imposes the
{Fronsdal equations} $\R_{\nu_1\ldots \nu_s}=0$
{and expresses the generalized HS Weyl tensors in terms of
derivatives of the Fronsdal fields}.

\section{Nonlinear higher-spin theory}

In this section we briefly summarize the construction of nonlinear
HS equations. The reader not interested in technical details is advised
to go directly to Section \ref{pro}.

\subsection{Idea of construction}
The idea is to look for nonlinear HS field equations in the form of a
nonlinear deformation of the Central on-shell theorem.
 The first step is to replace the linearized
HS curvatures and covariant derivatives by the full non-Abelian ones:
$$
R(y,\overline{y}\mid X) =d\go(y,\overline{y}\mid X)+
 \go(y,\overline{y}\mid X)*\go (y,\overline{y}\mid X)\,,
$$
$$
\tilde{D} C (y,\overline{y}\mid X) = dC (y,\overline{y}\mid X)
+\go (y,\overline{y}\mid X)*C (y,\overline{y}\mid X)-
C (y,\overline{y}\mid X)*\go (y,-\overline{y}\mid X)\,
$$
trying to find a deformation of the form
$$
R(Y\mid X) =  \overline{H}^{\dga\pb}
\f{\p^2}{\p \overline{y}^{\dga} \p \overline{y}^{\dgb}}\
{C}(0,\overline{y}\mid X) + H^{\ga\gb}
\f{\p^2}{\p {y}^{\ga} \p {y}^{\gb}}\
{C}(y,0\mid X) +\ldots\,,
$$
$$
\tilde{D} C(Y\mid X) +
  \ldots = 0\,,
$$
where further nonlinear corrections have to be determined from the formal consistency
of the HS equations. Having the form of the generalized Bianchi identities, the
consistency of the HS equations also guarantees their gauge invariance.

{Field equations of such a form are called {unfolded} which means that
all dynamical fields are differential forms and that the exterior derivative
of any field is expressed via the exterior product of the fields themselves.}
As discussed in some more detail in Section \ref{Unfolded Dynamics},
this form of dynamical equations is useful in many respects.

{Being possible in a few first orders, the straightforward construction of the
nonlinear deformation quickly gets complicated.}
{The trick is to find a larger algebra}
{$g^\prime$} {such that an appropriate substitution}
$$
 \go \to W=\go + \go C +\go
C^2+\ldots$$ { into} $W\in g^\prime$
{reconstructs nonlinear equations
via the flatness condition}
$$ dW+W\wedge W=0\,.
$$
{The problem is to find appropriate restrictions on} $W$ {that
reconstruct the nonlinear HS equations in all orders}.

{This is achieved via the doubling of spinors}
$$
\go(Y|X)\longrightarrow W(Z;Y|X)\,,\qquad
C(Y|X)\longrightarrow B(Z;Y|X)
$$
{accompanied by the equations that determine the dependence on the
additional spinorial variables} $Z^A$ {in terms of the ``initial data"}
$$
\go(Y|X)=W(0;Y|X)\q
C(Y|X)= B(0;Y|X)\,,
$$
where $\go(Y|X)$ and $C(Y|X)$ are the HS fields of the
Central on-shell theorem.
To rewrite the evolution along the additional variables $Z^A$
covariantly, it is useful to introduce {a connection
 $S(Z,Y|X)=dZ^A S_A$ in the $Z^A$-space.

\subsection{HS star product}

Nonlinear HS field equations are formulated in terms of the
specific star product
\be
\label{star2}
(f*g)(Z;Y)=\frac{1}{(2\pi)^{4}}
\int d^{4} U\,d^{4} V \exp{[iU^A V^B C_{AB}]}\, f(Z+U;Y+U)
g(Z-V;Y+V) \,,
\ee
where
$C_{AB}=(\gvep_{\ga\gb}, \bar \gvep_{\dga\dgb})$
is the $4d$ charge conjugation matrix and
$ U^A $, $ V^B $ are real integration variables. The
normalization is such that 1 is a unit element of the star-product
algebra, \ie $f*1 = 1*f =f\,.$ The star product
(\ref{star2}) is associative and provides a particular
realization of the Weyl algebra since
\be
[Y_A,Y_B]_*=-[Z_A,Z_B ]_*=2iC_{AB}\qquad
[Y_A,Z_B]_*=0\,.
\ee
 It results from the normal ordering
 with respect to the elements
\be
\label{oscrel}\nn
   b_A = \frac{1}{2i} (Y_A - Z_A )\,,\qquad
   a_A = \frac{1}{2} (Y_A + Z_A )\,,
\ee
which satisfy
\be
\label{com a}\nn
   [a_A, a_B]_*=[b_A, b_B]_* =0 \,,\quad
   [a_A, b_B]_* =C_{AB} \,
\ee
and can be interpreted as creation and annihilation operators.
In fact, the star product (\ref{star2})
 describes the normal ordering with respect to the oscillators $a_A$ and
  $b_A$ as
is most evident from the following consequences of (\ref{star2}):
\be\nn
\label{pr a}
   b_A * f(b, a) =
      b_A f(b, a) \,,\qquad
   f(b, a) * a_A =
      f(b, a)  a_A \,.
\ee

An important property of the star product (\ref{star2}) is that it
admits the inner Klein operator
\be\nn
\Upsilon = \exp i Z_A Y^A \,
\ee
which behaves as $(-1)^{N} ,$ where $N$ is the spinor number
operator. One can easily see that
\renewcommand{\U}{\Upsilon}
\be\nn
\U *\U =1,
\ee
\be
\label{[UF]}\nn
\U *f(Z;Y)=f(-Z;-Y)*\U
\ee
and
\be\nn
\label{Uf}
(\U *f)(Z;Y)=\exp{i Z_A Y^A }\, f(Y;Z) \,.
\ee

With respect to the decomposition of Majorana spinors  into
 two-component spinors,  $Y_A=(y_\ga,\by_\dga)$,
$\by_\dga = (y_\ga)^\dagger$,
the left and right inner Klein operators
\be
\label{kk4}
\kappa =\exp i z_\ga y^\ga\,,\qquad
\bar \kappa =\exp i \bar{z}_\dga \bar{y}^\dga\,
\ee
 act analogously on the undotted and dotted spinors, respectively,
\be\nn
\label{[uf]}
\kappa *f(z,\bar{z};y,\bar{y})=f(-z,\bar{z};-y,\bar{y})*\kappa\,,\quad
\bar{\kappa} *f(z,\bar{z};y,\bar{y})=f(z,-\bar{z};y,-\bar{y})*\bar{\kappa}\,,
\ee
\be\nn
\kappa *\kappa =\bar \kappa *\bar \kappa =1\q \kappa *\bar\kappa  = \bar\kappa*\kappa\,.
\ee

\subsection{The full nonlinear system}
As shown in \cite{Vasiliev:1992av}, the
equations of motion of the  four-dimensional HS theory can be
 formulated in terms of the three types of fields
\be\nn
W= dX^\nu W_\nu (Z,Y;K|X)\q
S=dZ^A S_A (Z,Y;K|X)\q B(Z,Y;K|X)\,.
\ee
The fields $W$ and $S$ are, respectively, one-forms in the four-dimensional space-time
with the coordinates $X^\nu$
and spinor space
with the coordinates $Z_A$. The spinorial variables
$Z_A$ and $Y_A$ are commuting while $dZ_A$ are anticommuting
differentials
\be\nn
Z_A Z_B = Z_B Z_A\q Y_A Y_B = Y_B Y_A\q
Z_A Y_B = Y_B Z_A\q dZ_A dZ_B =- dZ_B dZ_A\,.
\ee
$dZ_A$ commute with $Z_B$ and $Y_B$ but anticommute with the anticommuting
space-time differentials $dX^\nu$
\be\nn
dX_\nu dX_\mu = - dX_\mu dX_\nu\q dZ_A dX_\nu = - dX_\nu dZ_A\,.
\ee

 $K$ denotes a pair of Klein operators $K=(k,\bar k)$ that obey the relations
\be
\label{k2}
k^2={\bar k}^2 = 1\q k \bar k = \bar k k \,,
\ee
\be
\label{kw}
k w_\ga = - w_\ga k \q \bar k w_\ga = w_\ga \bar k\q
 k \bar w_\dga =  \bar w_\dga  k \q \bar k \bar w_\dga = -\bar w_\dga \bar k
\ee
 for $w_\ga = (dz_\ga, z_\ga, y_\ga )$, ${\bar w}_\dga =(d\bar z_\dga, \bar z_\dga,
 \bar y_\dga)$. The important difference between $(k,\bar k)$ and
 $(\kappa,\bar\kappa)$ (\ref{kk4}) is that the former anticommute with the respective differentials
 $dz^\ga$ and $d\bar z^\dga$ while the latter commute.

The system of nonlinear HS equations in $d=4$ reads as \cite{Vasiliev:1992av}
\be
\label{dW}
dW=W*W\,,
\ee
\be
\label{dB}
dB=W*B-B*W\,,\qquad
\ee
\be
\label{dS}
dS=W*S-S*W\,,
\ee
\be
\label{SB}
S*B=B*S\,,
\ee
\be
\label{SS}
S*S= -i (dZ^A dZ_A + dz^\ga dz_\ga  F_*(B)* k\kappa +
d\bar z^\dga d\bar
z_\dga \bar F_*(B) *\bar k \bar \kappa)
\,.
\ee
$F_*(B) $ in Eq.~(\ref{SS}) is some star-product function of the field $B$.
The simplest case of the linear functions
\be\nn
\label{etaB}
F_*(B)=\eta B \q \bar F_* (B) = \bar\eta B\,,
\ee
where $\eta$ is some phase factor (its absolute value can be absorbed into
a redefinition of $B$), leads to the class of pairwise nonequivalent nonlinear HS
theories. The  cases of $\eta=1$ and $\eta =\exp{\f{i\pi}{2}}$ are particularly
interesting, corresponding
to the so called $A$ and $B$ HS models. These two cases are distinguished
by the property that they  respect parity \cite{Sezgin:2003pt}.

Expanding all fields in powers of $k$ and $\bar k$ we obtain for $U=W,S,B$
\be\nn
U(Z,Y;K|X)=\sum^1_{i,j=0}k^i \bar k^j U_{ij}(Z,Y|X)\,.
\ee
Since the relations (\ref{k2}) and (\ref{kw}) are invariant under
the reflections $k\to -k$ and $\bar k \to -\bar k$, and taking into account that
the r.h.s. of Eq.~(\ref{SS}) contains $k$ and $\bar k$ explicitly, it follows
that the system (\ref{dW})-(\ref{SS}) is invariant under the following
involutive map
\bee
\tau (W(Y,Z;K|X)) = \ls&&W(Y,Z;-K|X)\q \tau (S(Y,Z;K|X)) = S(Y,Z;-K|X)\q\nn\\\nn
&&\tau (B(Y,Z;K|X)) = -B(Y,Z;-K|X)\,.
\eee
 As a result, the full system of fields decomposes into $\tau$-even
 and $\tau$-odd fields. Clearly, the $\tau$-even fields form a subsystem
 of the full system while the $\tau$-odd fields can be consistently
 truncated away. This truncation is applied in most of applications.
 The dynamical role of the $\tau$-even and $\tau$-odd  fields is different.

The $\tau$-even fields we call dynamical since they describe massless fields
of various spins. These are $W^{dyn}_{ii}$, $S^{dyn}_{ii}$ and
$B^{dyn}_{i\,1-i}$. Each of them appears in two copies because $i=1,2$.
As shown in \cite{Konshtein:1988yg}, this doubling is inevitable in presence
of fermions.

Each member of the infinite set of the $\tau$-odd fields describes
at most a finite number of degrees of freedom.
To stress that they carry no local degrees of
freedom, they were called auxiliary in \cite{Vasiliev:1992av}.
It is also appropriate to call
them {\it moduli} fields since the finite number of degrees of freedom
carried by each of these fields can be interpreted as a kind of coupling constants
of the theory. In particular, this was demonstrated in \cite{Prokushkin:1998bq}
 where it was
shown that the mass parameter of the matter fields in the $3d$ HS theory
results from a non-zero vacuum value of one of the moduli fields.
The moduli fields include
$W^{mod}_{i\,1-i}$, $S^{mod}_{i\,1-i}$ and
$B^{mod}_{ii}$.  Truncating away the moduli fields greatly
reduces the moduli space of the theory. In particular, the moduli responsible
for the massive boundary deformation can be argued to belong to this sector.

{The perturbative analysis} performed around the following
{vacuum solution}
$$B_0 =0\,,\qquad S_0 =dZ^A Z_A\,,\qquad
 W_0 = \half \go_0^{\mu\nu} (X) Y_\mu Y_{\nu}\,,
 $$
where $W_0$ obeys
$$
dW_0 +W_0 \star{}W_0 =0
$$
so that $\go_0^{\mu\nu}(X)$ describes the  $AdS_4$,
{reproduces  the Central on-shell theorem in the first-order \cite{Vasiliev:1992av}.} This means that the nonlinear system
(\ref{dW})-(\ref{SS}) indeed provides a nonlinear deformation of the free
equations of massless fields of all spins.
{Note that the specific form of the star product (\ref{star2}) is
crucial for this analysis.}

The HS equations exhibit {manifest gauge invariance under the gauge
transformations} $$ \delta W = d \gvep +[W\,,\gvep]_* \q
\delta S = [S\,,\gvep]_* \q \delta B = [B\,,\gvep]_*
\,,\qquad
\gvep=\gvep (Z;Y;K|X)\,. $$
{The nonlinear HS equations are formally consistent and regular: perturbatively, there are
no divergences due to star products of the non-polynomial elements resulting from
the inner Klein operators $\kappa$ and $\bar \kappa$} \cite{Vasiliev:1990vu}.

\subsection{Properties of HS interactions}
\label{pro}
Let us briefly  discuss some of the most important properties of the nonlinear HS equations.

{First of all, HS interactions contain higher derivatives. This property is closely
related to nonanaliticity  of the  HS interactions in} the cosmological constant $\Lambda=-\rho^{-2}$
which appears in the dimensionless combination $\rho \p$ where $\rho$ is the $AdS$
radius while $\p$ denotes the space-time derivative.
{This has the effect that background HS gauge fields contribute to the
higher-derivative terms in the evolution equations. As a result, the evolution
is determined mostly  by the HS
fields rather than by the metric. This provides the realization of the
anticipated property that
Riemannian geometry is not an appropriate tool in the HS theory.}

{In the HS theory,  HS fields source lower-spin fields in particular via}
the $\go*\go$-like {terms}. Other way around,
 lower-spin fields source HS fields via the $C^2$ {terms}.
{In particular, gravity sources the HS fields and vise-versa}.
Among other things this implies that the Einstein gravity cannot be obtained as
a consistent truncation of the HS theory.

{A remarkable feature of the HS  equations is that their nontrivial
part is only represented by Eq.~(\ref{SS}) which does not contain
the space-time derivative} $d$. This suggests that not only Riemannian
geometry but even usual coordinates do not play a fundamental role
in the system. In fact, this is a general property of unfolded dynamical
equations the particular case of which is represented by the nonlinear
equations (\ref{dW})-(\ref{SS}).

\section{Unfolded Dynamics}
\label{Unfolded Dynamics}
\subsection{General setup}
{The unfolded form of dynamical equations provides a covariant generalization
of the first-order form of  differential equations}
$$
\dot{q}^i(t) =\varphi^i (q(t))\,,
$$
which is convenient in many respects. In particular,
initial values can be given in terms of the values of variables
$q^i (t_0 )$ at any given point $t_0$. As a result, in the first-order
formulation, the number of degrees of freedom equals to the number of
dynamical variables.

{Unfolded dynamics is a multidimensional generalization achieved via the
 replacement of the time derivative by the de Rham derivative}
$$\f{\p}{\p t} \to d= \theta^\nu \p_\nu$$
and the dynamical variables $q^i$ by a set of differential forms
$$
 q^i(t)\rightarrow W^\Omega (\theta,x)=\theta^{\nu_1}\ldots \theta^{\nu_p}
W^\Omega_{\nu_1 \ldots \nu_p} (x) $$
to reformulate a system of partial differential equations in the
first-order covariant form
\be
\label{unf}
dW^\Omega (\theta,x)=G^\Omega (W(\theta,x))\,.\qquad
\ee
Here
$G^\Omega (W)$ {are some functions of the ``supercoordinates"}
$W^\Omega$
$$
G^\Omega (W) = \sum_n f^\Omega{}_{\Lambda_1\ldots
\Lambda_n}W^{\Lambda_1}  \ldots W^{\Lambda_n}\,.
$$
Since $d^2 =0$, at
$d>1$ {the functions $G^\Lambda (W)$ cannot be arbitrary but have to obey
the  compatibility conditions}
\be
\label{cc}
 G^\Lambda (W)\f{\p
G^\Omega (W)} {\p W^\Lambda}  \equiv 0\,.
\ee
(Recall that all products of the differential forms $W(\theta,x)$ are the wedge
products due to anticommutativity of  $\theta^\nu$.)
Let us stress that these are conditions on the functions $G^\Lambda (W)$ rather than on
$W$.

The idea of the unfolded formulation was put forward in the paper
\cite{Vasiliev:1990vu} where it was realized that the full system of
nonlinear equations can be searched in the form  (\ref{unf}) as a deformation
of the Central on-shell theorem.

As a consequence of the compatibility
conditions (\ref{cc}) the system (\ref{unf}) is manifestly invariant under the gauge
transformation}
$$ \delta W^\Omega = d \varepsilon^\Omega +\varepsilon^\Lambda
\frac{\p G^\Omega (W) }{\p W^\Lambda}\,,
$$
 {where the gauge parameter} $\varepsilon^\Omega (x) $ {is a}
$(p_\Omega -1)$-form if $W^\Omega$ is a $p_\Omega$-{form}. {Strictly
speaking, this is true for the class of {\it universal} unfolded systems in which
the compatibility conditions (\ref{cc}) hold independently of the dimension  $d$ of
space-time, \ie (\ref{cc}) should be
true disregarding the fact that any $(d+1)$-form is zero. Let us stress that
all unfolded systems,
which appear in HS theories including those considered in these lectures, are
universal.}

The unfolded formulation can be applied to the description of
invariant functionals of the system in question.  Here it is useful to
distinguish between
the off-shell and on-shell unfolded dynamical systems.

As demonstrated in
Section \ref{massunf}, most of the relations contained in
unfolded equations impose constraints expressing some new fields in terms of
derivatives of the old ones.  In the
\emph{off-shell} case the unfolded equations just express all
fields in terms of derivatives of some ground fields, imposing no differential
restrictions on the latter. In the scalar-field example of Section \ref{massunf},
to make the system off-shell one should relax the tracelessness condition in
(\ref{scal}). In this case, the pattern of the unfolded system (\ref{un0})
is given by the set of constraints (\ref{cons}) which express the higher tensors
$C_{a_1\ldots a_n}(x)$ via derivatives of the ground scalar field $C(x)$.
The \emph{on-shell} unfolded equations not only express all fields in terms of
derivatives of the ground fields, but also impose differential restrictions on the
latter. In the scalar-field example  this is the
Klein-Gordon equation (\ref{KG}).

As shown in \cite{Vasiliev:2005zu}, the variety of invariant functionals
associated with the unfolded equations (\ref{unf}) is described by the cohomology
of the operator
\be
\label{Q}
Q= G^\Omega \f{\p}{\p W^\Omega}\,,
\ee
which obeys
$$
Q^2=0
$$
as a consequence of (\ref{cc}). By virtue of (\ref{unf}), $Q$-closed $p$-form functions
$L_p(W)$ are $d$-closed, giving rise to the gauge invariant functionals
$$
S=\int_{\Sigma^p} L_p\,.
$$
In the off-shell case they can be used to construct invariant action functionals
while in the on-shell case they describe conserved charges. (For more detail and
examples see \cite{Vasiliev:2005zu}.) Also, in the on-shell case, $S$ can play a
r\'ole of the Hamilton-Jacobi action which becomes a functional of boundary
conditions in the context of holographic duality.\footnote{I am grateful to
Ioannis Papadimitriou for the
stimulating discussion of this point.}

\subsection{Properties}

The unfolded formulation of partial differential equations has a number of
remarkable properties.

\begin{itemize}
\item

First of all, it has general applicability:
 every system of partial differential equations can be
reformulated in the unfolded form.

\item

{Due to using the exterior algebra formalism, the system
is invariant under diffeomorphisms, being coordinate independent.}

\item

 Interactions can be understood as  nonlinear deformations of $G^\Omega(W)$.

\item

Degrees of freedom are represented by the subset of zero-forms $C^I(x_0)\in \{W^\Omega(x_0)\}$
{at any} {$x=x_0$}. This is analogous to the fact that $q^i(t_0)$
describe degrees of freedom in the first-order form of ordinary differential
equations.
The zero-forms $C^I(x_0)$ realize an {infinite-dimensional module dual to the space
of single-}{particle states of the system}. In the HS theory it is {realized  as
a space of functions of auxiliary variables like {$C(y,\bar y|x_0)$}. This space is
an analogue  of the phase space  in the Hamiltonian approach}.

\item

{It is worth to mention that the same property of the unfolded dynamics provides a
tool to control unitarity in presence of higher derivatives via the requirement that
the space of zero-forms like $C(y,\bar y )$ admits a positive-definite norm
preserved by the unfolded equations in question.}

\end{itemize}

The above list of remarkable properties of the unfolded formulation is far from
being complete. In particular, the unfolded formulation
 admits a nice interpretation in terms of Lie algebra
cohomology (for more detail see \cite{Vasiliev:2007yc}), $L_\infty$ algebra
\cite{Lada:1992wc}, $Q$-manifolds and many more
(for more detail see e.g.,  \cite{Vasiliev:2005zu,Bekaert:2005vh}
and references therein). The most striking feature of this formulation is however
that it makes it possible to describe one and the same dynamical system in
space-times of different  dimensions.

\section{Space-time metamorphoses}
\label{meta}
{Unfolded dynamics exhibits independence of the ``world-volume" space-time with
coordinates $x$.}
{Instead, geometry is encoded by} the functions $G^\Omega (W)$ in the ``target space"
of fields $W^\Omega$.
{Indeed, the universal unfolded equations make sense in any space-time independently of
a particular realization of the de Rham derivative $d$. For instance one can
extend space time by adding additional coordinates $z$}
$$
 dW^\Omega (x)=G^\Omega (W(x))\,,\quad x\rightarrow X=(x,z)\,,\quad
d_x\rightarrow d_X = d_x +d_z\,,\quad d_z =
dz^u\f{\p}{\p z^u}\,.
$$
The unfolded equations reconstruct the
$X${-dependence in terms of values of the fields} $W^\Omega(X_0)=W^\Omega(x_0,z_0)$
{at any} $X_0$. {Clearly, to take} $W^\Omega(x_0,z_0)$ in space $M_X$
{with coordinates} $X_0$ {is the same as to take} $W^\Omega(x_0)$ {in the space}
$M_x\subset M_X$  with coordinates  $x$.

The problem becomes most interesting
provided that there is a nontrivial vacuum connection along the additional
coordinates $z$. This is in particular the case of $AdS/CFT$ correspondence where
the conformal flat connection at the boundary is extended to the flat $AdS$
connection in the bulk with $z$ being a radial coordinate of the Poincar\'e type.

{Generally, the unfolding can be interpreted  as some sort of a covariant twistor
transform}
\bee\label{diapen}\nn
 \begin{picture}(200,80)( 0,26)
{\linethickness{.25mm}
\put(90,90){\vector( 1,-1){40}}%
\put(90,90){\vector( -1,-1){40}}%
\put(55,100)  {{    $C(Y|x)$}}
\put(6,32)  {{   $M(x)$}}
\put(128,32)  {{   $T(Y)\,.$}}
\put(35,70)  {{  \large $\eta$}}
\put(108,70)  {{  \large $\nu$}}
 }
\end{picture}
\eee
Here
 $W^\Omega(Y|x)$  {are functions on the
``correspondence space"} $C$ with  local coordinates $Y,x$.
{The space-time} $M$ {has local coordinates} $x$.
 {The twistor space}
 $T$ {has local coordinates} $Y$.

 {The unfolded equations reconstruct the dependence of $W^\Omega(Y|x)$ on $x$
 in terms of the function $W^\Omega(Y|x_0)$ on $T$  at some fixed $x_0$.
 The restriction of $W^\Omega(Y|x)$ or some its $Y$-derivatives to $Y=0$
 gives dynamical fields $\go(x)$ in $M$ which, in the on-shell case,
  solve their dynamical field equations.
 Hence, similarly to the Penrose transform (see \cite{Baston:1989vh}
 and references therein), unfolded equations  map functions
 on} $T$ {to solutions of the dynamical field equations in} $M$.

{In these terms, the holographic duality can be interpreted as the duality between
different space-times $M$ that can be associated with the same twistor space.}
This phenomenon has a number of interesting realizations.

\subsection{$AdS_4 /CFT_3$ HS holography}
The $AdS_4/CFT_3$ HS holography \cite{Klebanov:2002ja} relates the HS
gauge theory in $AdS_4$ to
the quantum theory of conformal currents in three dimensions. To see how it works,
let us first discuss the unfolded equations for free massless fields
and currents on the $3d$ boundary.

The unfolded  equations for conformal  massless fields in three dimensions are
\cite{Shaynkman:2001ip,Vasiliev:2001zy}
$$
(\f{\p}{\p x^{\ga\gb}} \pm i \f{\p^2}{\p y^\ga \p y^\gb} )
C_j^\pm(y|x)=0\q \ga,\gb=1,2\,,\quad j=1,\ldots {\mathcal N}\,.
$$
The equations for $3d$ conformal conserved currents have the form of rank-two equations
\cite{Gelfond:2003vh}
\be \label{xyy}\left\{
\,\f{\p}{\p x{}^{\ga\gb}}  -  \f{\p^2}{\p y^{(\ga}\p u^{\gb)}}
 \right\} J(u,\,y| x) =0\,.
 \ee
$J(u,\,y| x)$ {contains all $3d$ HS currents along with their
derivatives.

Elementary $3d$ conformal currents}, which are conformal {primaries},
{contain currents of all spins}
\be\nn
J(u,0|x)= \sum_{2s=0}^\infty u^{\ga_1}\ldots u^{\ga_{2s}}
J_{\ga_1\ldots \ga_{2s}}(x)\,,\quad \tilde
J(0,y|x)= \sum_{2s=0}^\infty y^{\ga_1}\ldots y^{\ga_{2s}}
\tilde J_{\ga_1\ldots \ga_{2s}}(x)\,
\ee
along with the additional scalar current
$$
J^{asym}(u,y|x) = u_\ga y^\ga J^{asym} (x)\,.
$$
Their conformal dimensions are
$$
\Delta J_{\ga_1\ldots \ga_{2s}}(x) = \Delta \tilde J_{\ga_1\ldots \ga_{2s}}(x)= s+1
\qquad
\Delta J^{asym}(x)=2\,.
$$

{The unfolded equations express all other components of $J(u,y|x)$ in terms of
derivatives of the primaries, also imposing the differential equations on the
primaries, which are just the conservation conditions
\be\nn
\f{\p}{\p x^{\ga\gb}}\f{\p^2}{\p u_\ga \p u_\gb} J(u,0|x) =0\q
\f{\p}{\p x^{\ga\gb}}\f{\p^2}{\p y_\ga \p y_\gb} \tilde J(0,y|x) =0\,
\ee
for all currents  except for the
scalar ones that do not obey any differential equations}.

The rank-two equation is obeyed by
\be\nn
J(u,\,y\,|x) =\sum_{i=1}^{\mathcal N}
C^-_{i}( {u}+y|x)\, C^+_{i}( y-u|x)\,.
\ee
This simple formula gives the explicit realization of the HS conformal
conserved currents in terms of bilinear combinations of  derivatives of
free massless fields in three dimensions.

{Generally, the rank-two fields and, hence conserved currents, can be interpreted as
bi-local fields in the twistor space.} In this respect they are somewhat
analogous to space-time bi-local fields also used for the description
of currents (see e.g \cite{Jevicki:2012fh,Nikolov:2001iz} and references
therein).

To relate $3d$ currents to $4d$ massless fields it remains to extend the
$3d$ current equation to the $4d$ {massless equations. This is easy to
achieve in the
 unfolded dynamics via the extension of the $3d$ coordinates} $x^{\ga\gb}$  to
the $4d$ coordinates $X^{\ga\dgb}$, extending the $3d$ equations to
\be
\label{Xyy}
\left ( \f{\p}{\p X^{\ga \dga}} + \f{\p^2}{\p y^\ga \p \bar y^\dgb} \right )
C(y,\bar y|X) =0\,.
\ee
{These are just the free unfolded equations for} $4d$  {massless fields of all
spins in Minkowski space, \ie at} $\Lambda=0$.

The analysis in $AdS_4$, which is also simple, is performed analogously.
In this case,
$
x^{\ga\gb} = \half ( X^{\ga \gb} + X^{\gb \ga})
$
{are boundary coordinates}, while
 $z^{-1} = X^{\ga\gb}\epsilon_{\ga\gb}$ is the {radial coordinate}.
 (For more detail see \cite{Vasiliev:2012vf}.)
 {At the non-linear level, the full HS theory in} $AdS_4$ {turns out to be
  equivalent to the theory  of } $3d$ {currents of all spins interacting
  through conformal HS gauge fields \cite{Vasiliev:2012vf}.}

\subsection{$sp(8)$ invariant setup}
Another example of the application of unfolded dynamics is related to the
$sp(8)$
extension of conformal symmetry in the theory of massless fields in four
dimensions. As was shown by Fronsdal \cite{F}, the tower of all
$4d$ {massless fields is} $sp(8)$ {symmetric}.
The $sp(8)$ symmetry extends conformal symmetry $su(2,2)\subset sp(8)$
that acts on every massless fields. The generators in $sp(8)/su(2,2)$
mix fields of different spins in the tower of massless fields of all spins
$0\leq s<\infty$.

Indeed, equations (\ref{dc}), that describe gauge invariant combinations
of massless fields, are covariant constancy conditions for 0-forms $C(y,\bar y|x)$
valued in the space of functions of spinor variables $y_\ga$ and $\bar y_\dga$.
Hence, symmetries of these equations contain $sp(8)$ realized by bilinears (\ref{bilin})
with indices $\ga$ taking $4$ values.

Fronsdal has shown that the space-time  $\M_4$ appropriate for
geometric realization of $Sp(8)$  is ten-dimensional {with local coordinates}
$X^{AB}=X^{BA}$, where $A=(\ga,\dga)=1,2,3,4$.
Applying the construction of Section 10, it is easy to derive the equations
for massless fields in $\M_4$.

\subsubsection{From four to ten}
Indeed, unfolded $4d$ massless equations can be easily
uplifted to $\M_4$ as follows \cite{Vasiliev:2001zy}:
\be
\label{un8}
dX^{AB}(\f{\p}{\p X^{AB}} +\f{\p^2}{\p Y^A \p Y^B} )C(Y|X)=0\q A,B=1,\ldots 4\,,
\ee
where, for the sake of simplicity, the massless equations are presented in
the Cartesian-like
coordinates. Note that to obtain the proper $\lambda\to 0$ limit from Eqs.~(\ref{dc}),
it is necessary to rescale the spinor variables
\be
y_\ga\to\lambda^{\half} y_\ga\q \bar y_\dga\to\lambda^{\half} \bar y_\dga
\ee
before taking the limit.

If the indices $A,B$  take just two values,  equations (\ref{un8})
describe $3d$ {massless fields} invariant under
$Sp(4)$ {which is the} $3d$ {conformal group} \cite{Shaynkman:2001ip,Vasiliev:2001zy}.

{By the general argument in the beginning of  Section \ref{meta},
equations (\ref{un8}) describe the same dynamics as
the original massless field equations in $4d$ Minkowski space because
they consist of the usual $4d$ equations (\ref{Xyy}) for the coordinates
$X^{\ga\dgb}$ supplemented with the
equations describing the evolution along the additional
spinning coordinates $X^{\ga\gb}$ and $X^{\dga\dgb}$.
The key question is what are independent dynamical variables in} $\M_4$?
From (\ref{un8}) it is clear that these are the fields $C(0|X)$ and $Y^A C_A (0|X)$.
Indeed, all other components of $C(Y|X)$ are expressed by Eq.~(\ref{un8})
via $X$-derivatives of $C(0|X)$ and $Y^A C_A (0|X)$.
It turns out that $C(0|X)$ {describes all} $4d$ massless fields of {integer spins}
while  $C_A (0|X)$ {describes all} $4d$ massless fields of {half-integer spins}.
So, $C(0|X)$ and $C_A (0|X)$ serve as certain hyperfields for the HS multiplets.

{The nontrivial field equations in $\M_4$ are} \cite{Vasiliev:2001zy}
\be
\label{10b}
\left (\f{\p^2}{\p X^{AB}\p X^{CD}}
-\f{\p^2}{\p X^{CB}\p X^{AD}}\right )C(X)=0
\ee
for bosons and
\be
\label{10f}
 \left (\f{\p}{\p X^{AB}}C_C(X)
-\f{\p}{\p X^{CB}}C_A(X)\right )=0
\ee
for fermions. These equations are interesting in many respects. First of all,
they are overdetermined. This is what  makes it possible to describe the
four-dimensional massless fields by virtue of differential equations in the
ten-dimensional space $\M_4$. Another interesting feature is that
equations (\ref{10b}) and (\ref{10f}) contain no index contraction and
hence no metric tensor.

\subsubsection{From ten to four}

It is instructive to see how the usual space-time picture re-appears
from the ten-dimensional one.
Remarkably, in this setup, the conventional four-dimensional picture
results from the identification of a
concept of local event simultaneously with the metric tensor.
Referring for more detail of the derivation to the original paper
\cite{Vasiliev:2001dc}, we just summarize the
final results.

{Time in} {$\M_M$} is a parameter $t$ along a time-like direction in
$\M_4$ represented by any positive-definite matrix $T^{AB}$
$$
X^{AB} = T^{AB}t\,.
$$

 Usual space in $\M_M$ is identified with the space of local events
 at a given time. Coordinates of the space of local events $x^n$ are required to
 have the property that the differential equations in question admit ``initial data"
 localized at any point of pace-time, \ie represented by the
 $\delta$-functions $\delta(x^n - x_0^n)$ with various $x^n_0$. Since the
 system of equations in question is overdetermined, the analysis of this issue is
 not quite trivial. The final result is  \cite{Vasiliev:2001dc}  that,
  for Eqs.~(\ref{10b}),(\ref{10f}), the space of local events in
 $\M_4$   is represented by a Clifford algebra with
$$
X^{AB} = x^n \gamma_n^{A}{}_C T^{BC}
$$
formed by matrices $\gamma_n^{A}{}_B$ that obey
\be
\{\gamma_n \,,\gamma_m\} = 2 g_{nm}\,,
\ee
where $g_{nm}$ is the spatial metric tensor of $R^3$.

Thus, the three-dimensional space of the $4d$ Minkowski space
appears as the space $R^3$ of local events. In this analysis,
the metric tensor appears just after the  identification
of coordinates that parametrize local events with the generators of the
Clifford algebra. In a certain sense, this construction is opposite
to the original Dirac's construction where the $\gamma$-matrices were introduced
as a square root of the metric tensor. Here,
the metric tensor appears from the definition of the $\gamma$-matrices that
represent local events.

Analogous analysis can be performed in some other dimensions.
In particular in \cite{Bandos:1999qf,Vasiliev:2001dc,Bandos:2005mb}
it was shown that equations (\ref{un8})
at $M=2,4,8,16$ describe free conformal fields of all spins in
$d=3,4,6,10$.

It should be noted that different $sp(2M)$-symmetric
 field equations in the same space $\M_M$,
like e.g. the higher-rank equations of \cite{Gelfond:2003vh}, have
spaces of local events of different dimensions. The resulting picture is somewhat
analogous to the brane picture in String Theory allowing the co-existence
of objects of different dimensions in the same space. The difference is
however that the ``HS branes"  in the $sp(2M)$ setup are not localized
as a particular surface embedded into $\M_M$. Instead, different choices of
a representative surface is a matter of the gauge choice. This example gives
another
manifestation of the general property that higher symmetries may
affect such fundamental concepts as local event and space-time dimension.

\section{HS theory and quantum mechanics}

Classical HS theory has several interesting links with quantum mechanics.

One is that {unfolded dynamics in the spinor (twistor) formulation
distinguishes between positive and negative frequencies}
\be
\label{meq}
\Big (\f{\p}{\p X^{AB}} \pm i \f{\p^2}{\p Y^A \p Y^B} \Big )
C^\pm(Y|X)=0\,.
\ee
Indeed, since {the time parameter}
$
t=\f{1}{M} X^{AB} T_{AB}$ {is associated with any positive-definite}  $T_{AB}$,
the sign in the exponential
$$
 C(X)= C^+(X) +C^-(X)\q  C^\pm(X) = \int \! d^M \xi
c^\pm (\xi) \exp \pm i\xi_A\xi_B X^{AB}
$$
is associated with the positive and negative frequencies.
Hence, the unfolded equations for massless fields in $\M_M$ effectively
quantize the model.

Another is  {the holographic duality between relativistic HS theory and  nonrelativistic
quantum mechanics}. To this end, consider the reduction of Eq.~(\ref{meq}) to the time arrow setting
  $X^{AB}=\delta^{AB}t$. The pullback of Eq.~(\ref{meq}) to the time axis gives
\be
\label{sch}
i \f{\p}{\p t} C^{\pm}(Y|t) =\pm \f{\p^2}{\p Y^A \p Y^B} \delta^{AB}
C^\pm(Y|t) \,.
\ee
We observe that this equation has the form of the  non-relativistic Schrodinger
equation for a free particle in the  space with coordinates $Y^A$.
Indeed, its right-hand side acquires the form of Laplacian in the variables
$Y^A$ while $C^\pm$ {play a role  of} $\psi$ { and} $\bar \psi$.

By the general argument of the beginning of this section, the two systems are equivalent,
\ie the relativistic HS theory in the $X$-space is equivalent to the nonrelativistic
theory in the twistor space. In particular, this equivalence manifests itself
in the equivalence of their symmetry algebra. As demonstrated in
\cite{Valenzuela:2009gu,Bekaert:2011qd},
the symmetry algebra of the Schrodinger equation is just the HS algebra
of Section \ref{Conformal HS}.

The Schrodinger equation (\ref{sch}) has zero potential. An interesting question is what
are dual HS theories for one or another nonzero potential. In the case of harmonic
potential the answer is known \cite{Vasiliev:2012vf}. The HS equations in
$AdS$ {and} $dS$ space-times are dual to the quantum-mechanical models with the
proper and upside down {harmonic potentials, respectively.
(Not surprisingly, the $dS$ geometry corresponds
to the unstable quantum mechanics.)

{Since the HS theory has a potential to unify gravity with quantum mechanics,
one can speculate that it may be able to shed light on the both ingredients.}
{Since full HS theory is nonlinear, its identification with quantum mechanics
at the linearized level may suggest that, at  ultrahigh energies, the HS theory
may affect the fundamentals of quantum mechanics itself, making it nonlinear
with the gravitationally small coupling constant!}

\section{To String Theory via Multiparticle Symmetry}

Properties of the HS theory are to large extent determined by the properties of
the HS algebra. It has been long anticipated that the HS theory should be related
somehow to String Theory. To materialize this idea it is most important
to find a HS algebra rich enough to underly the full fledged String Theory.
Recently it was conjectured \cite{Vasiliev:2012tv} that such a symmetry can be
associated with
a multiparticle symmetry that acts on all multiparticle states of the HS theory.

Mathematically, this symmetry algebra can be defined as the Lie algebra
associated with the universal enveloping algebra of the HS algebra
of Section \ref{Conformal HS}.
It has a number of features that make it promising as a candidate for a
string-like extension of the HS theory. In particular,
it contains the original HS algebra  {as a subalgebra}. Acting on all
multiparticle states of HS theory it has enough room for
mixed symmetry fields which appear in String Theory.

If this idea will indeed work, it will allow to interpret String
Theory as a theory of bound states of the HS theory in striking analogy
with the conjecture of \cite{Chang:2012kt}.

\section{Summary and Conclusion}

{The HS gauge theories contain gravity along with infinite towers of other fields
with various spins including ordinary matter fields. An interesting feature
of any HS model is that it always contains a scalar field associated with
graviton, which  carries no internal indices. It is tempting to
speculate that this scalar may play a role in cosmology and, specifically,
for inflation.}

The {HS theory contains non-minimal higher-derivative interactions that make
it a kind of a nonlocal theory with unusual properties. In particular,
many of the standard tools of GR based on Riemannian geometry may not
be applicable to the HS theory as a  consequence of the
fact that the HS symmetry transforms a spin-two field to HS fields. In practice,
this implies that in HS theories one has to be careful with the conventional interpretation
of physical phenomena in terms of the metric tensor.} In particular, this should be taken
into account in the analysis of black hole physics in the framework of HS theory.

{The HS gauge theories exist in any dimension
\cite{Vasiliev:2003ev}. However, the HS theories
available so far are  analogues of pure supergravity with no matter
multiplets included. This makes it difficult to analyze the important issue of
spontaneous breakdown of the HS symmetry which is necessary to introduce a mass
scale analogous to the string tension.}
In fact, it can be argued that, to achieve a spontaneous breakdown of the
 HS symmetry, the string-like extension of the HS theory is needed. It was
 recently conjectured
\cite{Vasiliev:2012tv} that such an
extension can be  provided by a multiparticle theory to be identified with
the quantum HS theory and String Theory.

{Another exciting feature of the HS theory is that it exhibits a
remarkable interplay between classical and quantum physics. This suggests
that the further analysis may shed some light on both  gravity and quantum mechanics
at transplanckian  energies which is the regime to be described by the HS theory.

HS theories not only have interesting holographic duals but also, being
formulated in terms of unfolded dynamics approach, can shed
light on the very origin of holographic duality. It can be argued \cite{Vasiliev:2012vf}
that the holographic duality links such models in space-times of different
dimensions, that have equivalent form of their unfolded equations.

There are many  important directions of the research  of HS theories I had no
chance to touch in these lectures.

One of the most interesting is the construction of exact solutions
of HS equations. Most of exact solutions available so far, one way or another
result from the solution of $3d$ HS theory obtained in \cite{Prokushkin:1998bq}.
There are two main types of exact black-hole type solutions of the
nonlinear HS equations available in the literature.  The first one
is represented by the flat connections associated with the BTZ-like black holes
in the $3d$ HS theory (see \cite{Ammon:2012wc} and references therein;
the interpretation of the usual BTZ black hole
\cite{Banados:1992wn} as a solution of
the HS equations was given in \cite{Didenko:2006zd}).
The second type includes black hole solutions in $AdS_4$ with the nonzero
curvature tensor  \cite{Didenko:2009td,Iazeolla:2011cb}.
Some other solutions were considered e.g. in
\cite{Sezgin:2005pv,Sezgin:2005hf,Iazeolla:2007wt}.
Analysis of their properties in the context of the HS holographic duality
and beyond is an important direction of the current research.

Among other activities we should mention analysis of the action
principle in HS theory at the cubic level (see, e.g.,
\cite{Metsaev:2005ar,Alkalaev:2010af,Vasilev:2011xf,Buchbinder:2012iz,Metsaev:2012uy,Joung:2012hz,Boulanger:2012dx})
and beyond \cite{Boulanger:2011dd,Boulanger:2012bj} as well as the
 further progress in understanding HS holography (see e.g.,
\cite{Maldacena:2011jn,Maldacena:2012sf,Didenko:2012vh,Afshar:2013vka,Giombi:2013fka,Metsaev:2013wza,
Tseytlin:2013jya,
Giombi:2014iua,Giombi:2014yra}) including holographic RG flows
\cite{Douglas:2010rc,Sachs:2013pca,Leigh:2014tza}
and conformal correlators of HS currents
\cite{Giombi:2012ms,Zhiboedov:2012bm,Colombo:2012jx,Didenko:2012tv,Gelfond:2013xt,
Stanev:2013qra,Florakis:2014kfa}.

Since in this short review it is hard even to list all important research
directions in the HS theory, we refer the reader to other reviews
\cite{Vasiliev:1995dn,Vasiliev:1999ba,Bekaert:2005vh,Bekaert:2010hw,Sagnotti:2011qp,
Gaberdiel:2012uj,Giombi:2012ms,Ammon:2012wc,Didenko:2014dwa} as well as to
the  contribution of Ricardo Troncoso to this workshop \cite{Perez:2014pya},
where more detail and references on various aspects of the HS theory can be
found.

\section* {Acknowledgments}

The author is grateful to Olga Gelfond for useful comments on the
manuscript and to the organizers of  the 7th Aegean workshop on
non-Einstein theories of gravity on Paros for the warm atmosphere and
hospitality. This research was supported in part by RFBR Grant
No 14-02-01172.

\end{document}